\documentclass[11pt]{article}
\usepackage{moriond,epsfig}
\usepackage{amsmath}

\def\be{\begin{equation}}
\def\ee{\end{equation}}
\def\bea{\begin{eqnarray}}
\def\eea{\end{eqnarray}}

\begin{document}
\vspace*{4cm}
\title{ALL-ORDER CORRECTIONS TO HIGGS BOSON PRODUCTION IN ASSOCIATION WITH JETS}

\author{JEPPE R.~ANDERSEN}

\address{Theory Division, Physics Department, CERN, CH 1211 Geneva 23, Switzerland}

\maketitle\abstracts{ We present a new framework for calculating multi-jet
  observables. The framework is based on the factorisation of scattering
  amplitudes in the kinematical limit of large invariant mass between all
  particles.  We show that by constraining the analyticity of scattering
  amplitudes away from this limits, we get good agreement order by order with
  the full, fixed order perturbative calculation at the low orders where
  these are available, and therefore get firm predictions on the all-order
  behaviour. As an example, we study Higgs boson production through gluon
  fusion in association with at least two jets at the LHC.}
\vspace{-10mm}
\section{Introduction}
\label{sec:introduction}
Achieving the full potential of the LHC will challenge our understanding and
description of events with multiple jets. A detailed understanding and
description of the multi-jet predictions arising from the Standard Model (SM)
is necessary in order to fully disentangle this contribution from that which
might arise from much sought-after extensions of the Standard Model.

The true complication of each observed jet in terms of its constituent
hadrons can currently only be described within the context of a parton shower
and hadronisation model, as implemented in
e.g.~Ref.~\cite{Sjostrand:2007gs,Bahr:2008pv,Gleisberg:2008ta}. However,
while obtaining a good description of the \emph{structure} of each jet, the
underlying soft and collinear resummation often underestimates the
\emph{rate} and \emph{hardness} ($p_\perp$-spectrum) of multi-jet
samples\cite{Alwall:2007fs}. Matching
procedures\cite{Catani:2001cc,Lonnblad:1992tz,Mangano:2006rw} can ensure that
the description of a given process contains at least the full
tree-level. Virtual corrections (and the resulting weighting of samples with
varying jet multiplicity) are, however, estimated using only the Sudakov
factor from the shower, which arises from the requirement of unitarity of the
parton shower.

In this contribution, we will describe results obtained in an approach, which
sums perturbative corrections to the hard scattering matrix element to all
orders, but considers corrections of a different origin than that of the soft
and collinear logarithms of the parton shower. Instead of focusing on the
emission under small invariant masses, we will focus on the limit of hard,
wide-angle emission. The goal is to achieve an all-order exclusive
description of the leading radiative corrections (real and virtual) for the
formation of extra jets, and not be concerned with the description of the
internal structure of each jet, which we will leave for a later matching with
a parton shower.

\section{Building Blocks for All-Order Results}
\label{sec:building-blocks-all}
The approximation, which eventually allows us to construct an exclusive
(i.e.~differential in the momenta of all particles), all-order resummation of
the hard scattering matrix element, is based on keeping only the leading
contribution to scattering amplitudes in the limit where the invariant mass
$s_{ij}$ between all particles is large. The QCD radiative corrections to the
basic $2\to2$ partonic process (or $2\to Wjj$\ldots) can then be calculated
in this limit. In terms of the rapidities $y_i$ and transverse momenta
$p_{\perp,i}$ of each particle, this \emph{Multi-Regge Limit} is written as
\begin{equation}
y_0\gg y_1\gg \ldots\gg y_{n+1};\quad |p_{i\perp}|\simeq |p_{i+1\perp}|,
\label{MRK}
\end{equation}
where obviously $y_0\gg y_1$ really means $y_0-y_1\to\infty$. In fact, the
leading contribution to the tree-level matrix elements of \emph{all}
scattering processes can be calculated in this strict limit (and sub-leading
contributions are suppressed by one power of $s_{ij}$ in the square of the
matrix element). This MRK limit of the matrix elements is reproduced by a set
of Feynman rules consisting of just one Feynman diagram for each rapidity
ordering of particles. This one Feynman diagram consists of a string of
gluon propagators connecting effective
vertices\cite{Fadin:2006bj,Bogdan:2006af} like e.g.~the non-local
\emph{Lipatov vertex}
\begin{align}
C^{\mu_i}(p_a,p_b,q_i,q_{i+1})&=\left[-(q_i+q_{i+1})^{\mu_i}-2\left(\frac{{\hat{s}}_{ai}}{\hat{s}_{ab}}
+\frac{\hat{t}_{i+1}}{\hat{s}_{bi}}\right)p_b^{\mu_i}\right.
\left.+2\left(\frac{\hat{s}_{bi}}{\hat{s}_{ab}}
+\frac{\hat{t}_i}{\hat{s}_{ai}}\right)p_a^{\mu_i}\right],
\label{lip1}
\end{align}
here described for the scattering process $p_a\ p_b\ \to p_0\cdots p_n$,
where $\hat{s}_{ai}=2p_a\cdot p_i$ etc., and $\hat{t}_i=q_i^2$ is the
propagator associated with the $i^{\text{th}}$ connecting gluon. In fact, the
leading virtual corrections can be parametrised to all orders by replacing
the $1/q_{i}^2\to 1/q_{i}^2\exp({\hat\alpha(q_i)(y_{i-1}-y_i)})$ in the
propagators.

So if these effective Feynman rules results in an approximation of the
scattering amplitude which reproduces the known MRK limit of the full
scattering amplitude, why not just use this limit instead of the (slightly)
more involved effective Feynman rules? The point is that we would like an
approximation for the \emph{inclusive} corrections to all orders,
i.e.~without having to require large rapidity separations between \emph{each
  and every} set of particles. By using the effective Feynman rules, we can
ensure that aside from reproducing the correct MRK limit, the amplitudes
fulfill certain requirements when applied away from the MRK limit (obviously,
any phase space point relevant to the LHC is specifically \emph{away} from
the exact MRK limit). Firstly, the amplitudes which arise from these
effective Feynman rules are gauge-invariant, i.e.~satisfies the Ward identity
$k.M=0$ \emph{exactly} for each gluon of any (on-shell) momentum $k$, not
just in the MRK limit. Secondly, the full kinematic dependence is kept in the
divergences arising from the propagators of the connecting $t$-channel gluons
(i.e.~there is no limit taken in the kinematic invariants arising); see
Ref.\cite{Andersen:2008gc,Andersen:2008ue} for more details.

By also capturing the leading (in $\log(s_{ij}/t_i)$) contribution to the
virtual corrections, it is straightforward to organise the cancellation of
the infra-red poles between the real and virtual corrections, see
e.g.~Ref.\cite{Andersen:2008gc}. The resulting amplitudes are sufficiently
simple that they can be evaluated to any (necessary) order in $\alpha_s$, and
the phase space integration is efficiently implemented following the
procedure of Ref.\cite{Andersen:2006sp}. The end result is an
\emph{inclusive} (in the sense of including the emission of any number of
gluons) calculation, which is \emph{exclusive} in the momenta of all
particles. Therefore, any analysis (jet-algorithm, etc.) can be implemented
on the output of the resummation.
\vspace{-4mm}
\section{Higgs Boson plus Multiple Jets}
\label{sec:higgs-boson-plus}
This resummation scheme was first applied to the process of Higgs Boson
production through gluon fusion in association with at least two
jets\cite{Andersen:2008gc,Andersen:2008ue}. This process is particularly
interesting when a large rapidity difference between two jets
occurs. Firstly, it allows for an extraction of the CP structure of the
ttH-coupling\cite{Klamke:2007cu}. Secondly, it is necessary to understand the
process in details, since it is a background to the extraction of the
coupling of the Higgs boson to Z/W in Higgs boson production through weak
boson fusion~\cite{Hankele:2006ma}. This process therefore naturally lends
itself to a treatment based on the phase-space assumptions of
Eq.~(\ref{MRK}).

\begin{figure}[tb]
  \centering
  \epsfig{width=.49\textwidth,file=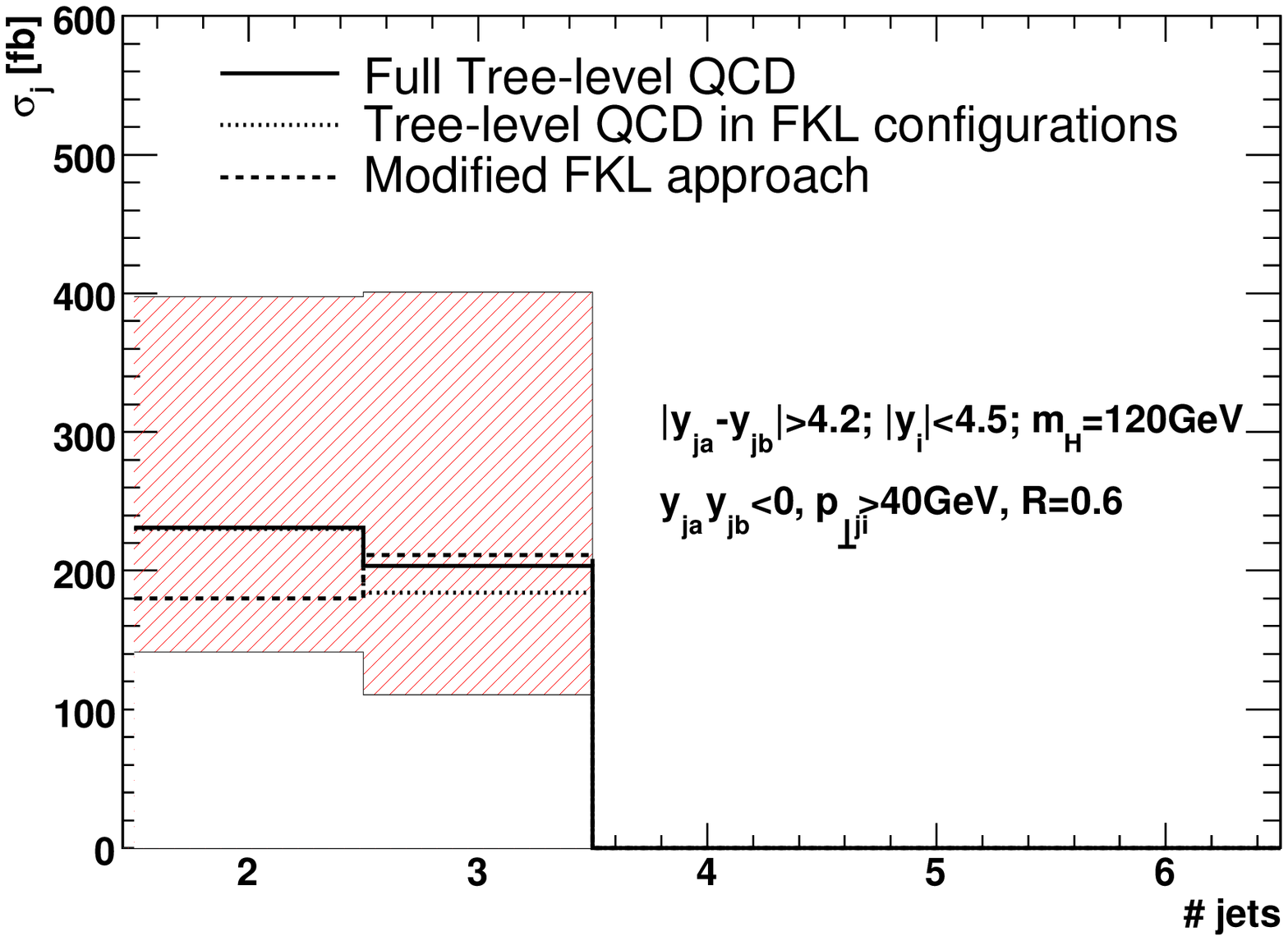}
  \epsfig{width=.49\textwidth,file=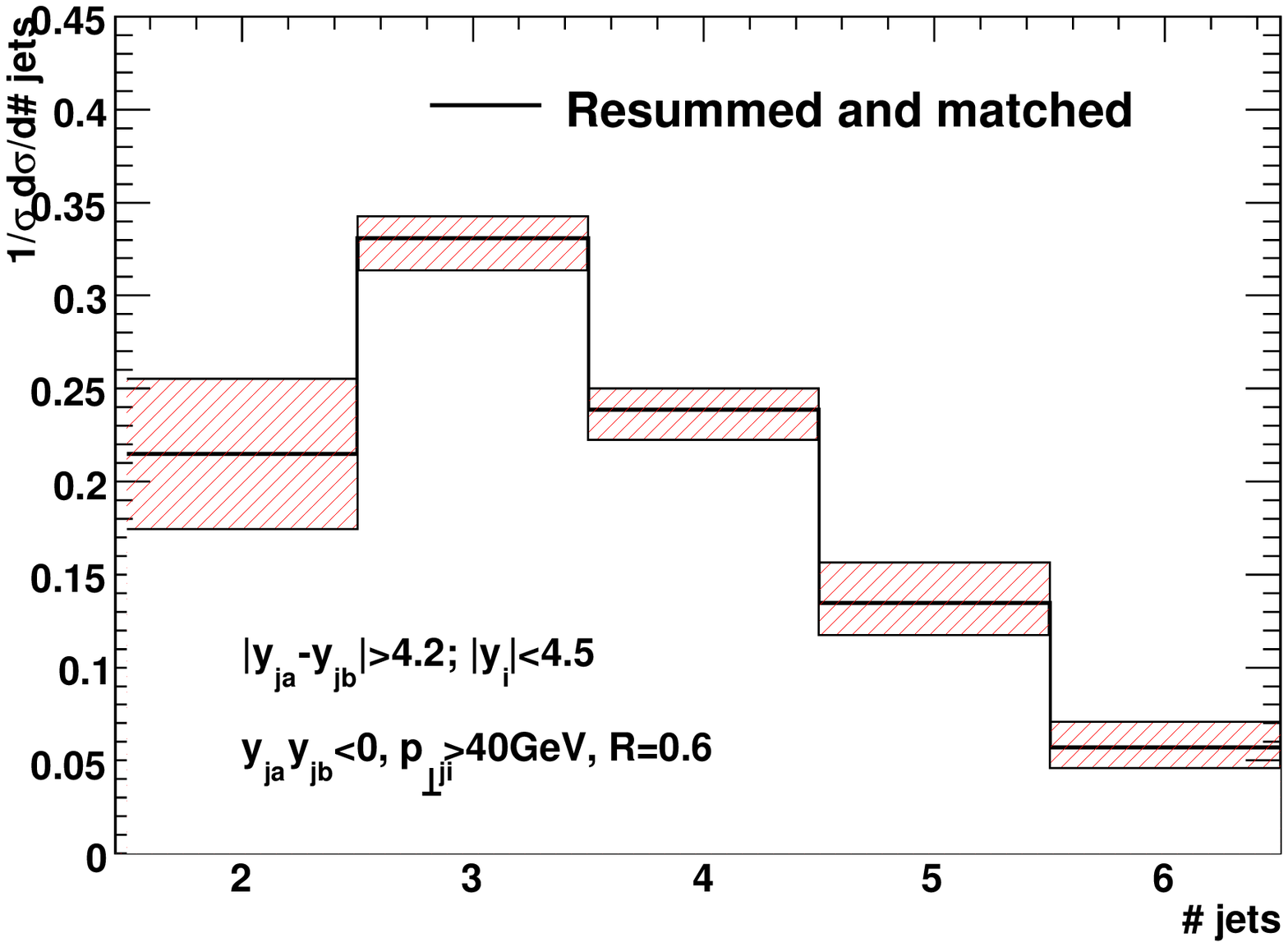 }
  \caption{\textbf{Left:} The cross section for hjj and hjjj obtained using
    MadGraph, compared with the result obtained using the effective Feynman
    rules, allowing all-order resummation. \textbf{Right:} The relative jet
    rates in the fully resummed and matched event sample.}
  \label{fig:fixed_comp_and_resum}
\end{figure}

After constructing a set of Feynman rules which are sufficiently \emph{simple} to
allow all-order results to be constructed, the next step for any resummation
programme should be to verify that the rules are also sufficiently
\emph{accurate} that whatever is summed will resemble the perturbative series
for whatever process is claimed to be resummed. On
Fig.~\ref{fig:fixed_comp_and_resum}(left) we compare the $\alpha_s^4$ and
$\alpha_s^5$ cross sections for hjj and hjjj (jets defined with kt-algorithm,
$p_\perp>40$GeV) production respectively, between
our approximation (allowing all-order results to be obtained) and the full
tree-level QCD results (obtained using MadGraph\cite{Alwall:2007st}), within
a standard set of \emph{weak-boson fusion}-cuts, and with a scale choice
equal to a Higgs mass of 120GeV. The red bands indicate the scale uncertainty.
It is clear that the approximation is sufficiently accurate in describing the
hard emission that it is worthwhile constructing a resummation based on
them. This holds true also for kinematic distributions, see
Ref.\cite{Andersen:2008gc}. Furthermore, in the final formulation the Higgs boson
plus two and three jet results are matched to full tree-level accuracy. On
Fig.~\ref{fig:fixed_comp_and_resum}(right), we show the relative contribution
from various exclusive jet states within the resummed and matched result for
inclusive Higgs boson production in association with at least two
jets. Again, the red bands indicate the scale uncertainty in the relative jet
rates in the resummed result. We see that within these cuts, the exclusive
two-jet rate will account for only roughly 17-25\% of the cross section for
Higgs boson production in association with two or more jets, with the higher
jet rates accounting for the rest.

\begin{figure}[htb]
  \centering
  \epsfig{width=.49\textwidth,file=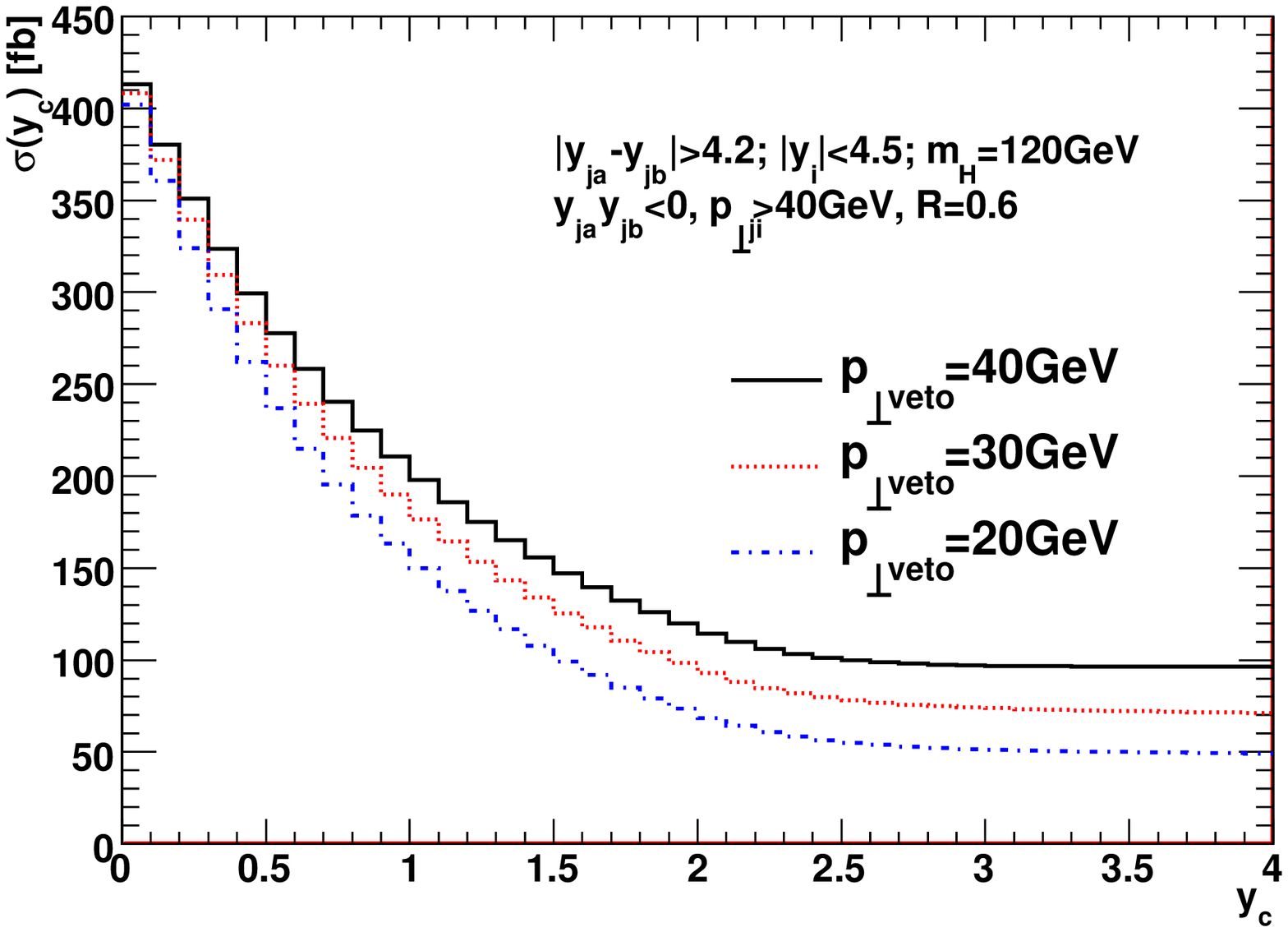}
  \caption{The cross section for Higgs boson production in association with
    at least two jets, as a function of the parameters of a veto on extra jet
  activity.\vspace{-8mm}
}
  \label{fig:jetveto}
\end{figure}

The inclusive (i.e.~containing virtual and real-emission corrections to all
orders) nature of the resummation allows one to study the dependence on the
details (like rapidity-range and transverse momentum cut) of a central jet
veto. Such jet vetos are intended to suppress the gluon-fusion contribution
to the hjj-channel. As an example, we show in Fig.~\ref{fig:jetveto} the
cross section in the resummed and matched calculation, when apart from the
cuts mentioned on the figure, a requirement is imposed that
\begin{equation}
\forall j\in\{\mbox{jets\ with\ } p_{j\perp}>p_{\perp,\mbox{veto}}\}\setminus\{a,b\}\ :\ \left|y_j-\frac{y_a+y_b}{2}\right|>y_c,
\label{eq:yc}
\end{equation}
where jets $a,b$ are the most forward/backward hard jet of transverse
momentum larger than 40GeV. For a transverse momentum cut of $40$GeV, the
result for $y_c\to\infty$ is obviously what would be called the exclusive
two-jet rate of the resummed and matched calculation. It is seen that with a
veto as hard as 40GeV on further jets, the cross section is reduced to about
100fb (with the scale choices made in Ref.\cite{Andersen:2008gc}), which is
less than half of the tree-level prediction for hjj within the same
cuts. This promises well for central jet vetos as a method of suppressing
gluon fusion contribution to hjj within the weak boson fusion cuts.
\section{Conclusions}
\label{sec:conclusions}
\vspace{-2mm}
We have very briefly described the ideas behind a resummation scheme, which
captures the effects of hard emission resulting in the formation of
observable jets, and discussed example analyses. A partonic event generator
based on this formalism for Higgs boson production in association with jets
can be downloaded \texttt{http://andersen.web.cern.ch/MJEV}.
\vspace{-5mm}
\section*{References}



\end{document}